\documentclass{emulateapj}
\usepackage{placeins}
\usepackage{epsfig,graphicx}
\usepackage{float}
\usepackage[pass]{geometry}
\restylefloat{table}
\def\lap{\lower.5ex\hbox{$\; \buildrel < \over \sim \;$}}
\def\gap{\lower.5ex\hbox{$\; \buildrel > \over \sim \;$}}

\def\ergcm2s{${\rm erg\ cm^{-2}\ s^{-1}}$}

\def\ergscm2s{${\rm erg\ cm^{-2}\  s^{-1}}$}

\def\cm-2{${\rm cm^{-2}}$}

\begin{document}

\title{A Global Star Forming Episode in M31 2-4 Gyr Ago}

\author{Benjamin F. Williams\altaffilmark{1},
Julianne J. Dalcanton\altaffilmark{1},
Andrew E. Dolphin\altaffilmark{2},
Daniel R. Weisz\altaffilmark{1,4},
Alexia R. Lewis\altaffilmark{1},
Dustin Lang\altaffilmark{3},
Eric F. Bell\altaffilmark{5},
Martha Boyer\altaffilmark{6},
Morgan Fouesneau\altaffilmark{7},
Karoline M. Gilbert\altaffilmark{8},
Antonela Monachesi\altaffilmark{9},
Evan Skillman\altaffilmark{10}
}

\altaffiltext{1}{Department of Astronomy, Box 351580, University of Washington, Seattle, WA 98195; ben@astro.washington.edu, jd@astro.washington.edu, dweisz@astro.washington.edu}
\altaffiltext{2}{Raytheon, 1151 E. Hermans Road, Tucson, AZ 85706; adolphin@raytheon.com}
\altaffiltext{3}{McWilliams Center for Cosmology, Department of Physics, Carnegie Mellon University, 5000 Forbes Ave., Pittsburgh, PA; dstn@cmu.edu}
\altaffiltext{4}{Hubble Fellow}
\altaffiltext{5}{Department of Astronomy, University of Michigan, 550
  Church St., Ann Arbor MI 48109; ericbell@umich.edu}
\altaffiltext{6}{Observational Cosmology Lab, Code 665, NASA Goddard Space Flight Center, Greenbelt, MD 20771 USA; martha.boyer@nasa.gov}
\altaffiltext{7}{MPIA, Heidelberg; fouesneau@mpia-hd.mpg.de}
\altaffiltext{8}{STScI; kgilbert@stsci.edu}
\altaffiltext{9}{MPA, Garching; antonela@mpa-garching.mpg.de}
\altaffiltext{10}{Department of Astronomy, University of Minnesota, 116 Church
St. SE, Minneapolis, MN 55455; skillman@astro.umn.edu}

\begin{abstract}

We have identified a major global enhancement of star formation in the
inner M31 disk that occurred between 2-4 Gyr ago, producing $\sim$60\%
of the stellar mass formed in the past 5 Gyr. The presence of this
episode in the inner disk was discovered by modeling the optical
resolved star color-magnitude diagrams of low extinction regions in
the main disk of M31 (3$<$R$<$20~kpc) as part of the Panchromatic
Hubble Andromeda Treasury.  This measurement confirms and extends
recent measurements of a widespread star formation enhancement of
similar age in the outer disk, suggesting that this burst was both
massive and global.  Following the galaxy-wide burst, the star
formation rate of M31 has significantly declined.  We briefly discuss
possible causes for these features of the M31 evolutionary history,
including interactions with M32, M33 and/or a merger.

\end{abstract}

\section{Introduction}

Perhaps the only external massive galaxy for which we can hope to
obtain detailed observational evolutionary constraints with the
current generation of technology is the Local Group large spiral
galaxy M31. Its mass \citep[1.4${\times}10^{12}$
  M$_{\odot}$;][]{watkins2010} and morphology
\citep[SA(s)b;][]{devaucouleurs1991} are similar to those of the
galaxies that dominate redshift surveys.  Its nearby distance
\citep[770 kpc;][]{mcconnachie2005}, low foreground extinction
\citep[$A_V$ of 0.17 mag;][]{schlafly2011}, and relatively favorable
inclination \citep[$\sim$77$^{\circ}$ not constant,][]{braun1991} make
it possible to study in great detail. All of these characteristics make
M31 an excellent specimen to further our understanding of the
formation and evolution of galaxies.

Even though M31 is so nearby, there are many challenges to unlocking
its evolutionary history.  First, its individual stars are not
resolved from the ground beyond the top few magnitudes of the stellar
luminosity function \citep{massey2006}, and even these stars are often
strongly affected by blending.  Second, it is highly inclined and
dusty \citep{dalcantondust}, complicating any interpretation of the
stellar populations of the disk itself.  Third, it is home to a
massive bulge component that dominates the stellar mass inside of
$\sim$1.5 kpc \citep[e.g.,][]{courteau2011,dorman2013}, further
complicating interpretations of observational measurements.  Fourth,
the halo structure and satellite galaxy distribution are both complex,
with the halo showing streams and multiple metallicity components
\citep{ibata2001,ferguson2002,mcconnachie2009,brown2003,kalirai2006a,kalirai2006b,gilbert2009,gilbert2014,ibata2014a}
and the satellite population containing a co-planar set of dwarfs
\citep{ibata2013}.

Although the evolution of M31 has been somewhat difficult to
determine, several resolved star studies have shown that the disk is
dominated by stars older than 1 Gyr
\citep{williams2002,bellazini2003}, and very deep resolved photometry
of the outer disk suggests the disk is somewhat younger than the halo
population \citep{brown2006}.  Further work on the outer disk has
suggested a widespread (25--90 kpc deprojected radii) star formation
episode at $\sim$2 Gyr ago that may correspond to an encounter with
M33 \citep{bernard2012,bernard2015}, which also shows a peak in that
age range \citep{williams2009}. This idea is consistent with the many
cold streams seen in the halo of M31, which are indicative of recent
tidal interactions and many merger events
\citep{ibata2001,ferguson2002,mcconnachie2009,ibata2014}.  In fact,
the halo of M31 appears so much more complex than that of the Galaxy,
that several have suggested that M31 could have undergone relatively
recent (2-6 Gyr ago) mergers
\citep{ibata2005,hammer2010,hammer2013,sadoun2014}.  Additionally, a
relatively coeval population of massive star clusters was found by
\citet{fusipecci2005} with an age of $\sim$2 Gyr, suggesting a
significant star formation event at that time.  Finally, very recent
analysis of the kinematics of the M31 disk stars show that their
velocity dispersion is significantly higher than that of the Galaxy,
suggesting a more active merger history \citep{dorman2015}.

In this paper, we make a new contribution to our understanding of the
evolution of M31 by measuring the detailed age distribution of the
stellar populations within the inner disk, showing how the burst seen
in the outer galaxy persists to small galactic radii.  By picking out
the stellar photometry of the dust-free regions of the PHAT survey, we
have been able to make reliable model fits to the optical color
magnitude diagrams for 9 regions, ranging from 3 to 20 kpc from the
galactic center (deprojected).  The results show a
globally-significant episode of star formation 2-4 Gyr ago, confirming
and magnifying the significance of the result of \citet{bernard2015}.
Section 2 briefly describes the data set used and our fitting
techniques, which are both described in detail elsewhere in the
literature.  Section 3 gives the results of our measurements, focusing
on the past 5 Gyr, and Section 4 discusses the possible
interpretations of the results, including interaction and merger
scenarios.

\section{Data}

%\subsection{The Panchromatic Hubble Andromeda Treasury}

The data we have used for this project is a subset of the photometry
from the Panchromatic Hubble Andromeda Treasury
\citep[PHAT;][]{dalcanton2012,williams2014}.  These data include
ultraviolet (UV)
through near infrared (NIR) resolved stellar photometry, as well as a sample of
artificial stars taken at a range of stellar densities.  The
artificial stars are sufficient to determine the uncertainties and
completeness as a function of wavelength and stellar density
\citep{williams2014}.

\subsection{Identifying Low-Extinction Regions}

Historically it has been challenging to model the stellar populations
of M31 inside of 20~kpc because of the high dust content which causes
significant differential reddening
\citep[e.g.][]{dalcanton2012,draine2014}.  To overcome this
difficulty, we take advantage of the large contiguous area of the PHAT
survey to find small regions containing very little extinction.

The photometry sample for our study was chosen using the extinction
maps of \citet{dalcantondust}, where dozens of 15$''{\times}15''$
(60pc~x~250pc, deprojected) dust-free regions were found across PHAT
footprint by measuring the width of the red giant branch in the NIR.
We selected 365 of these regions, and grouped them into 8 distinct
areas to provide a wide range of radius and azimuth: five areas along
the major axis, one along the minor axis, and three areas of large
deprojected radius.

\begin{figure}
\includegraphics[width=3.5in]{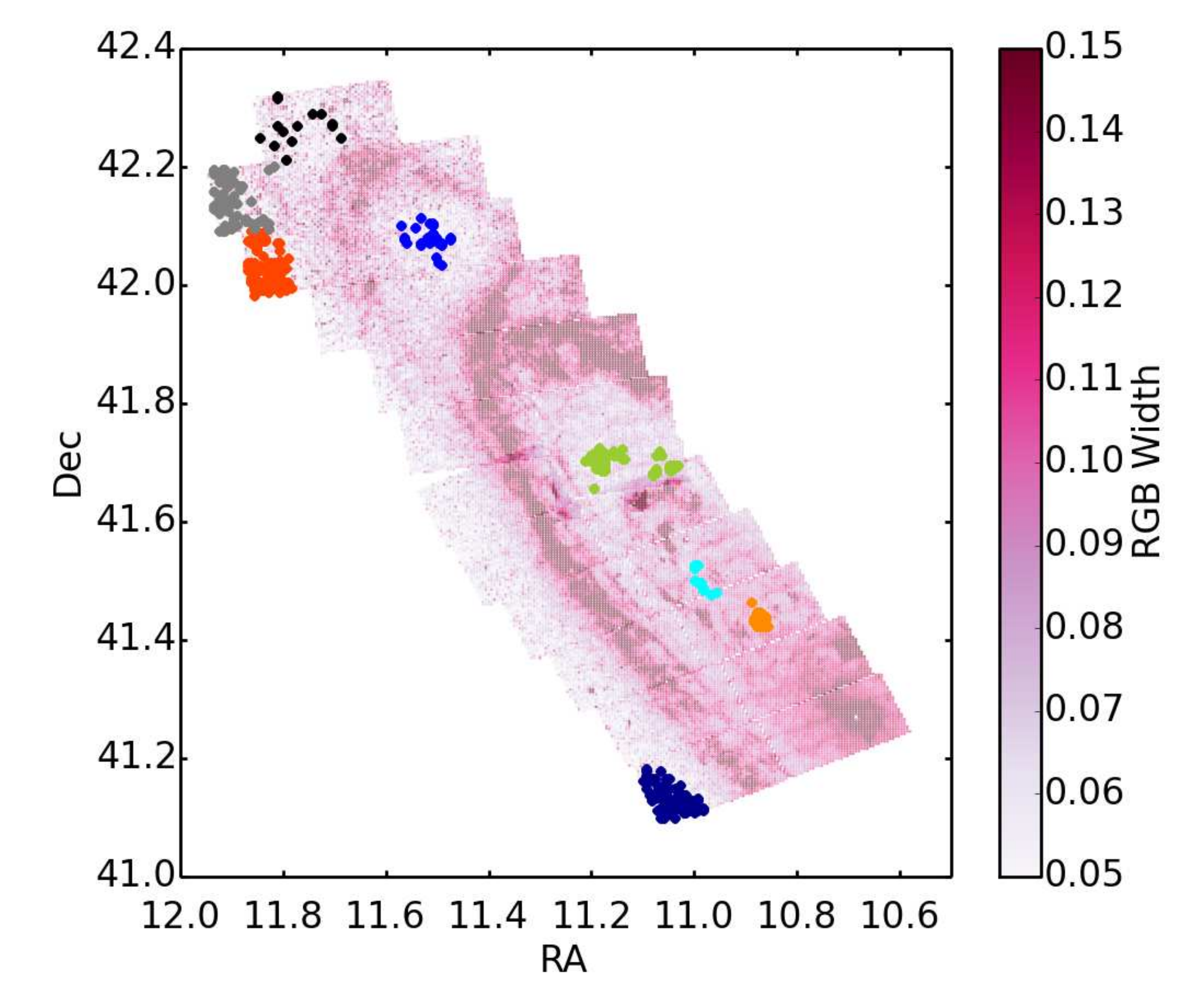}
\includegraphics[width=3.5in]{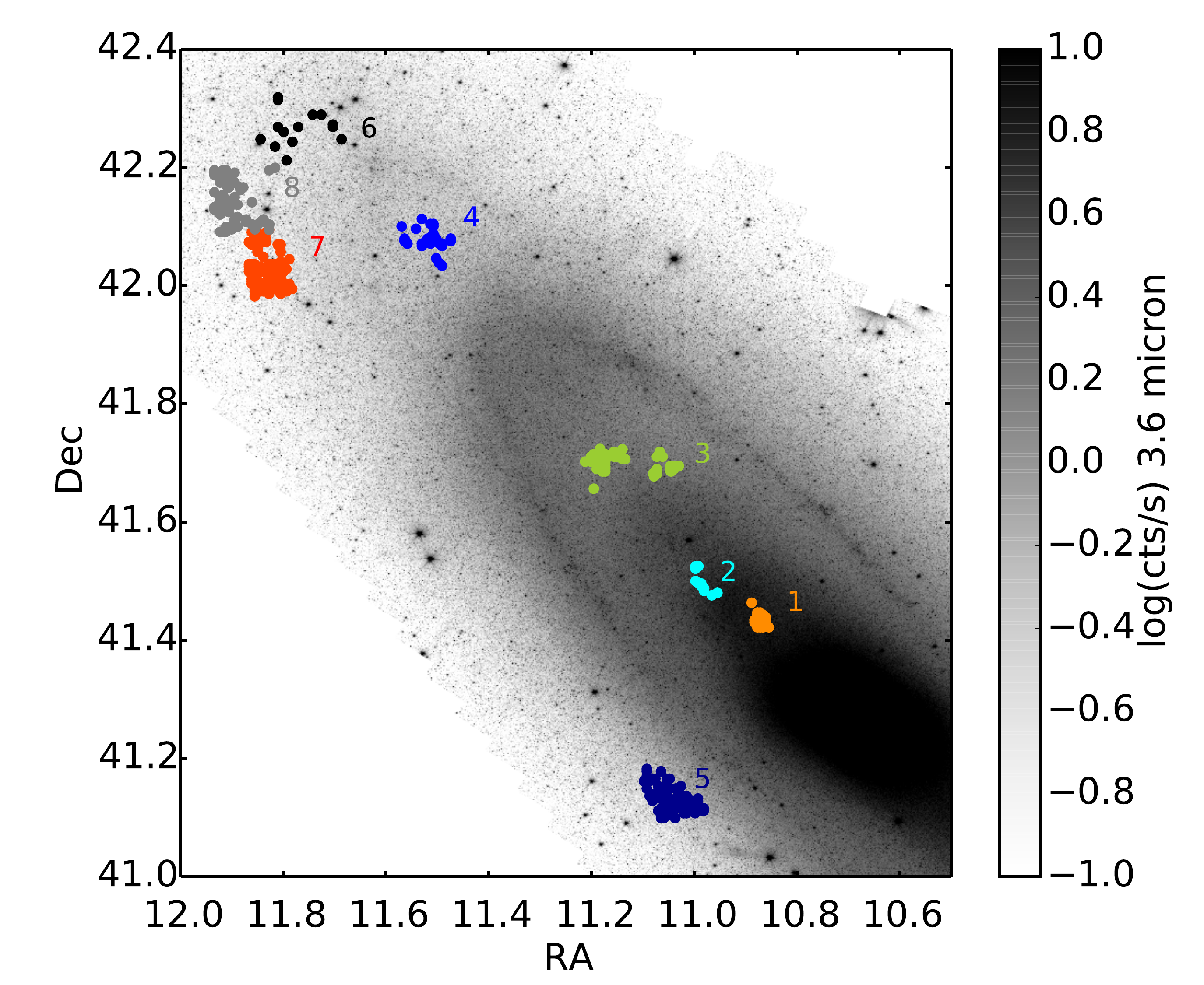}
\caption{{\it Left:} Locations in the PHAT footprint of our
  ``reddening-free'' samples are overplotted on a map of the width of
  the red giant branch as measured in the NIR photometry. This width is
  a proxy for the amount of dust present along the line of sight. The
  sample comes from the regions of narrowest width, and therefore
  containing very little dust.  Circles of the same color show how we
  grouped the samples to improve our CMD-fitting statistics.  {\it
    Right:} Locations in the PHAT footprint of our ``reddening-free''
  samples are overplotted on a {\it Spitzer} 3.6 $\mu$ image, showing
  the wide range of stellar densities covered by our samples.
  However, groups are confined to areas with little stellar density
  variation.  Numbers label each group with their region number in
  Table~\ref{locations_tab}.}
\label{locations}
\end{figure}

Figure~\ref{locations} shows our groups plotted on both a map of the
RGB width within the PHAT footprint as well as a 3.6 micron Spitzer
map of M31.  Table~\ref{locations_tab} provides the median, minimum,
and maximum deprojected radii, total area, total number of stars, and
best-fit differential reddening for each group.  The group numbers in
the table correspond to the labels on the figure.

\subsection{Measuring Star Formation Histories}

\begin{figure}
\includegraphics[width=6.5in]{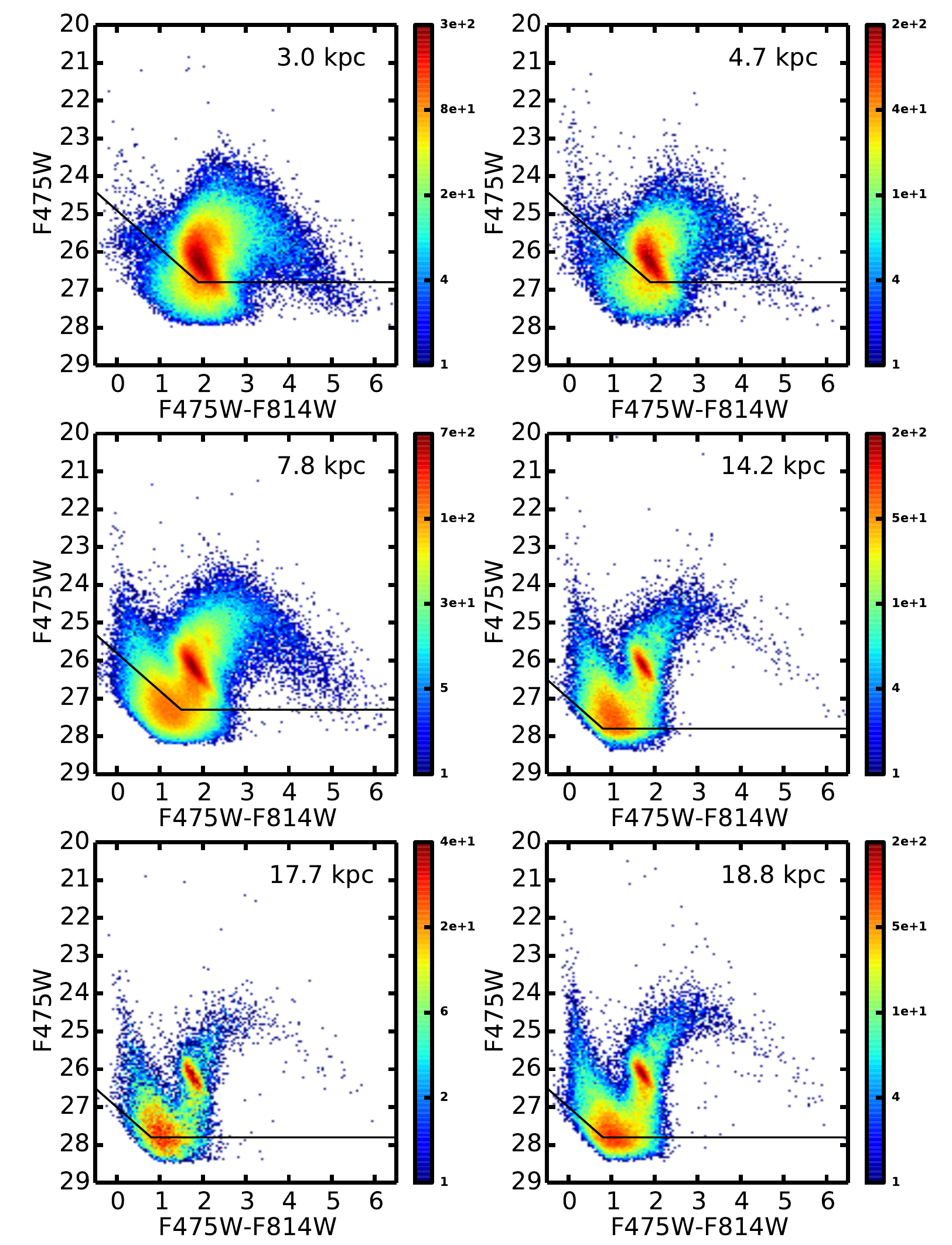}
\caption{The color-magnitude diagrams of 6 of our sample groups at
  different radii along the major axis.  Other regions are not shown
  for brevity. Color gives the number of stars per 0.05x0.05 mag CMD
  bin. All data above the black line was included in the fitting.  All
  show well-behaved red giant branches, with no indication of
  differential reddening of more than $\sim$0.5 mag.  Crowding
  significantly limits photometric precision and depth inside of 5
  kpc.}
\label{cmds}
\end{figure}

The color-magnitude diagrams of 6 of our 8 groups, which span the
range of photometric quality of our data, are shown in
Figure~\ref{cmds}.  While the red giant branch (RGB) of samples from
smaller radii are broader due to crowding effects \citep[see][for
  details]{williams2014}, these RGBs are narrow and well-defined.  In
addition, the red clump feature, visible at F814W$\sim$24.5 in all
CMDs sensitive to that depth, shows very little, if any, extension.
These CMDs are consistent with the \citet{dalcantondust} measurement
of these regions containing $<$0.6 mag of differential extinction. As
a result, we allowed up to 0.6 mag of differential extinction (d$A_V$)
in our model fits.  Including this small amount of differential
extinction improved the fits, and did not change the measured age
distributions beyond our measured uncertainties. However, it did
impact some of the details of the star formation histories (SFHs) at
intermediate ages in the 2 innermost groups.  These groups were
best-fit with d$A_V$=0.4-0.6 mag.  The exact choice of d$A_V$ had no
effect on our result within our uncertainties.  However, if no
differential extinction was included at these radii, the results
differed significantly, and were considerably less consistent with the
results at larger radii.

We chose these regions because of their lack of dust.  Therefore,
they are biased against containing recent star formation.  Thus, we
have confined our comparison across the disk to ages $>$500 Myr.
Ages younger than this are explored in detail in \citet{lewis2015} and
are highly structured spatially.  By 500 Myr, the structure is still
apparent, though less pronounced. In any case, this bias is
unavoidable for our present analysis, so we will take it into
consideration in any interpretations of our results.

Each of these groups has an appropriate set of artificial stars
measured from a region of similar stellar density.  To maximize our
age sensitivity, we fit the optical CMDs of each set of photometry and
artificial stars using the software package MATCH, using techniques
well documented in the literature
\citep{dolphin2002,gallart2005,dolphin2012,dolphin2013,weisz2014}.  In short, the CMD is
binned into a 2-D histogram, and this histogram is fitted by a linear
combination of model histograms from a fine grid of age and
metallicity.  The best-fit provides the most likely age and
metallicity distribution for the stars.  We then apply a Monte Carlo
technique for determining the range of SFHs that would provide
acceptable fits to the data.  The results of the Monte Carlo runs
yield random uncertainties for our measurements given the reliability
of the models \citep{dolphin2013}.  

To determine the significance of a particular feature in the
best-fitting SFH, one must account for systematic uncertainties that
come from deficiencies in the models or offsets between the models
and measurements.  The
technique for determining these uncertainties is detailed in
\citep{dolphin2012}.  Briefly, the photometry is fitted to the models
in 50 independent runs.  In each run, the models are shifted with
respect to the data by a random and small amount ($\pm$0.02 in log of
the effective temperature and $\pm$0.17 in bolometric magnitude).  The
results of the 50 runs are merged to calculate the variance of the
results in each time bin, which is adopted as the systematic
uncertainty.  These uncertainties were combined with the random
uncertainties to produce our total uncertainties on our age
distributions.

We fit our data using several different stellar evolution models.  We
adopted fits with the Padova models \citep{girardi2000,marigo2008} as our
fiducial SFHs; however, we show the nominal SFHs using the BASTI
\citep{pietrinferni2004} and PARSEC \citep{bressan2012} stellar
evolution libraries to give a sense of the effects that
different models have on age when fitting data of our varying depths.  

We also produced model CMDs of constant star formation rate over the
past 14 Gyr, convolved with the errors, completeness, and reddening
measured from the PHAT data of varying depths.  We fitted these using
the same techniques used to fit the observed CMDs.  As with the
extensive tests of such fitting in the literature
\citep[e.g.][]{dolphin2002}, in our tests, the measured SFHs fell
within the uncertainties of the input constant SFH at the full range
of photometric depths included in our data.

%To confirm our results, and test the reliability of the absolute age
%and duration of the last burst of star formation in M31, we also fit
%the NIR photometry of a few groups to check for consistency.  Although
%the NIR data are significantly shallower, they were useful in
%determining if our results were strongly biased due to any unknown
%model deficiencies in the optical.
\subsection{Comparisons with Deeper Photometry}

\begin{turnpage}
\begin{figure}
\hspace{1.0cm}\includegraphics[width=3.0in]{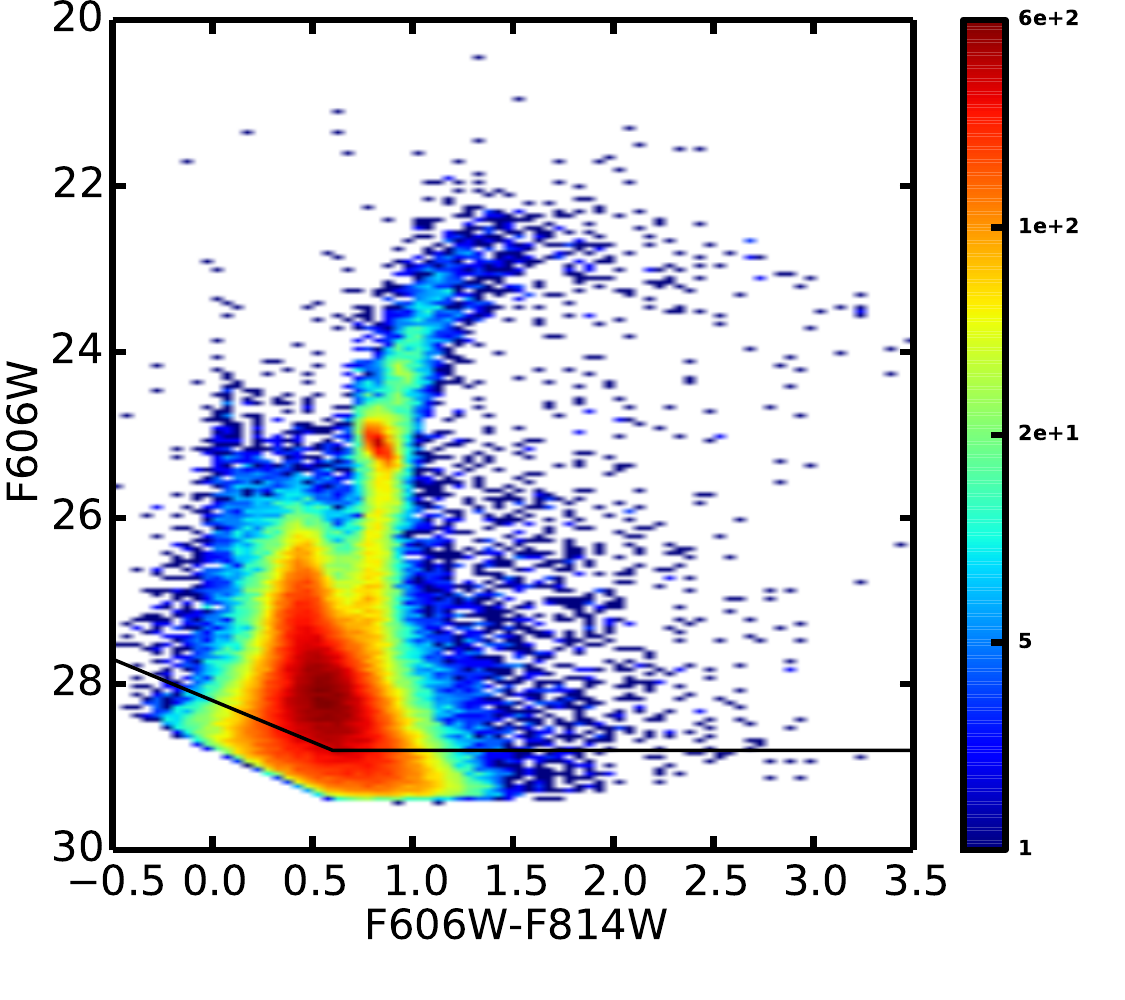}
\includegraphics[width=3.0in]{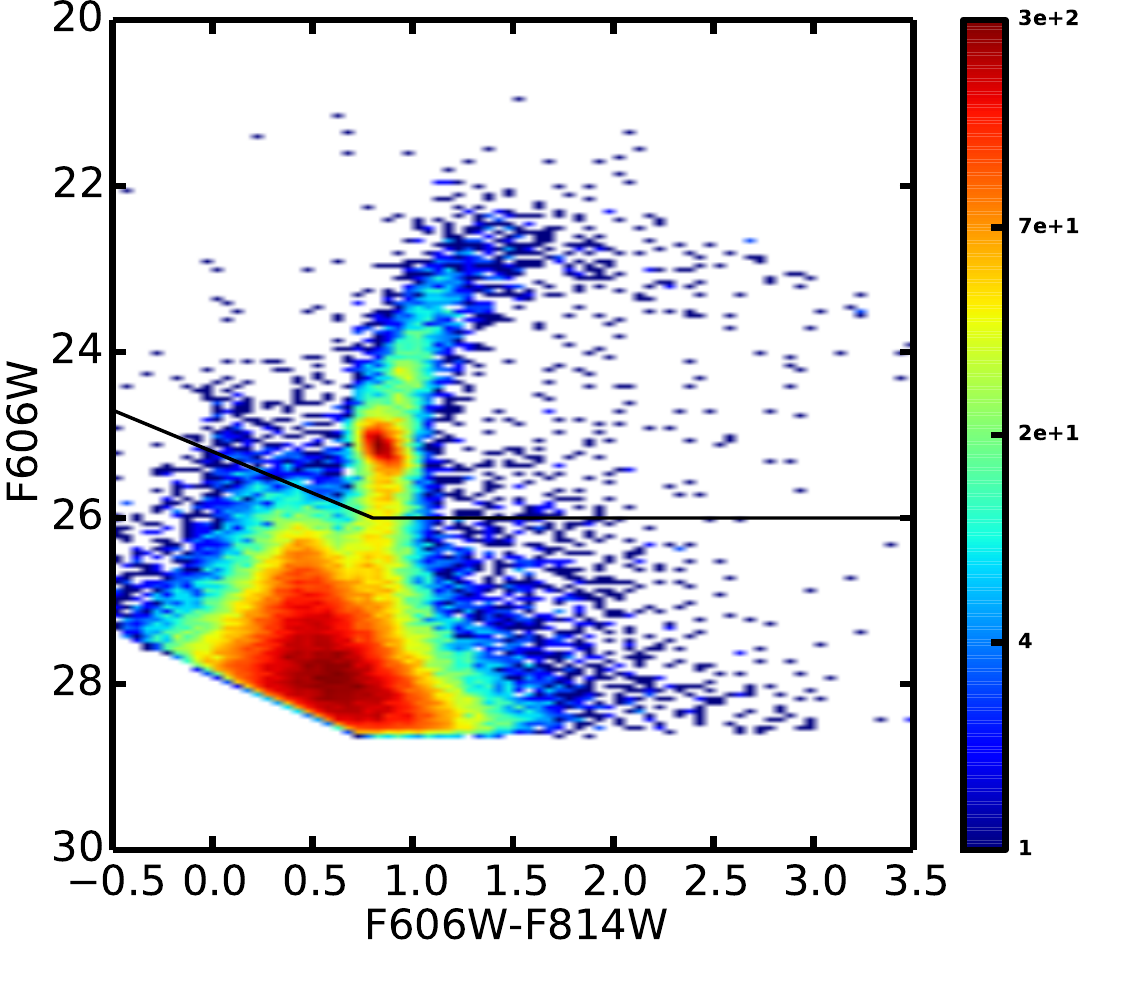}
\includegraphics[width=2.8in]{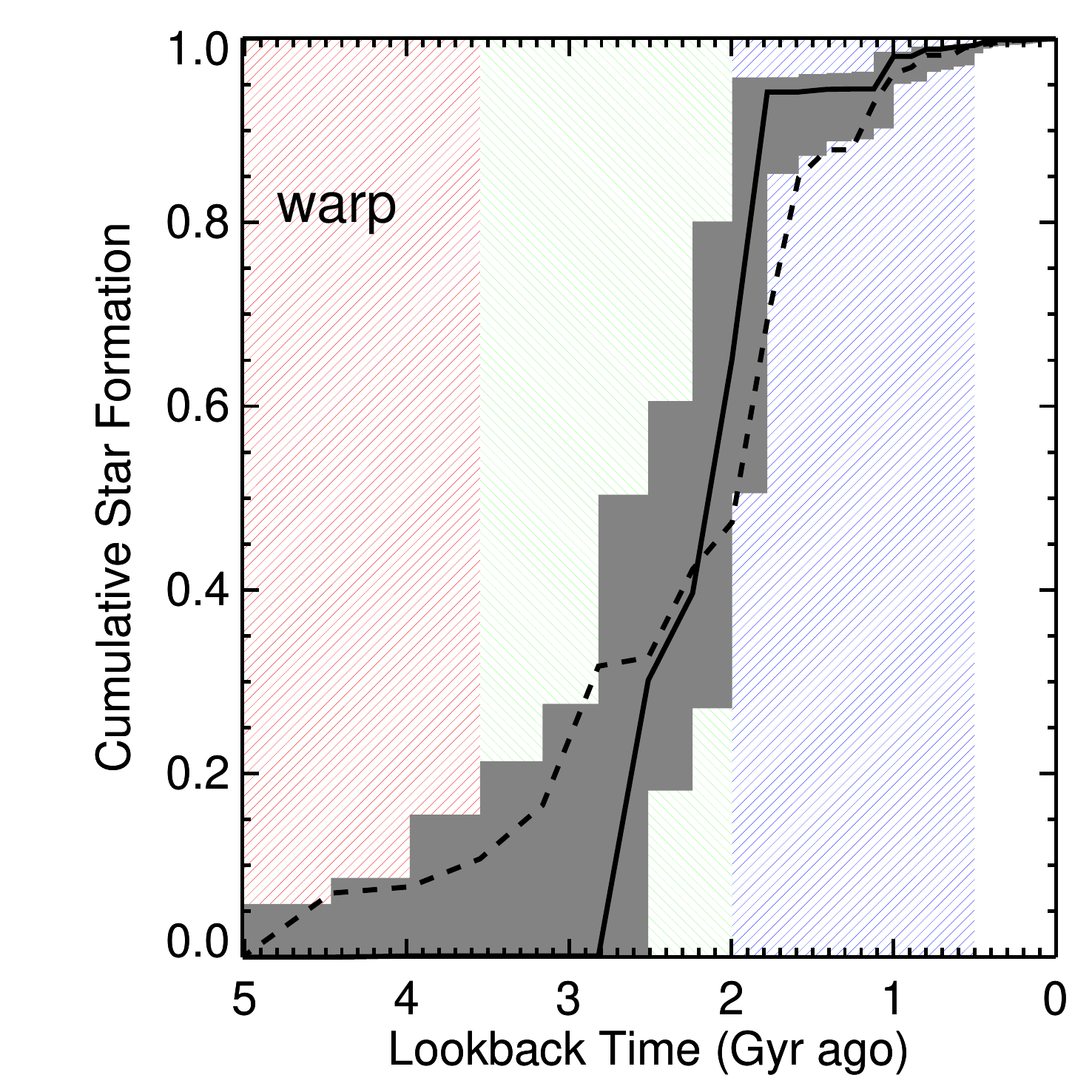}
\caption{Testing the effects of depth on the recovered star age
  distribution over the past 5 Gyr.  Although the CMDs all contain a
  significant fraction of stars older than 5 Gyr which were also
  measured by the SFH fitting, we leave the contribution from these
  stars out of the current analysis to focus on the epoch of interest.
  Thus all of the cumulative fractions are relative to only the subset
  of stars with ages $<$5 Gyr, not the total stellar population.  {\it
    Upper Left:} Color-magnitude diagram of our photometry from the
  warp field of \citet{bernard2012}, using their full set of HST
  exposures. We included all of the CMD above the black line in the
  fitting. {\it Upper Right:} Same as left, but from photometry
  performed using only 2 exposures in each band, of similar depth to
  the PHAT data.  We included all of the CMD above the black line in
  the fitting. {\it Lower Left:} The cumulative star formation history
  of the warp field from \citet{bernard2012}. Three colored stripes
  show the 3 epochs used to make the plots in Figure~\ref{profiles},
  in their respective colors.  The dashed line shows the best fit to
  the full-depth data, which is similar to their measurements.  The
  solid line shows the best fit to the shallow photometry, and the
  dark gray shaded area shows the absolute uncertainties associated
  with the fit to the shallow photometry. Both measurements clearly
  detect the burst at $\sim$2 Gyr, showing that even though the
  precision of the measurement from the shallow data is much worse,
  the absolute uncertainties are robust. {\it Lower Right:} The
  differential SFH of both fits.  The thin black histogram shows the
  best-fit SFH to the full-depth data with associated uncertainties,
  and the thick gray histogram shows the best-fit SFH to the shallow
  data with associated uncertainties.  Again, both measure a burst at
  $\sim$2 Gyr, although the shallower data finds a slightly older and
  less precise age than the full-depth data.}
\label{comp_bernard}
\end{figure}

\end{turnpage}

In addition to our new measurements, we tested our ability to recover
reliable SFHs using data of the depth and quality of the PHAT
photometry.  Our test data were taken from the HST archive using the
ACS field in the warp in the outer regions of the southern M31 disk,
previously analyzed in detail by \citet{bernard2012}.  This field has
very deep data and contains a clear feature at $\sim$2 Gyr in the
published age distribution.  We analyzed this field in two ways.
First, we ran the entire data set through our entire photometry and
CMD-fitting technique with the Padova models to ensure that we could
reproduce the \citet{bernard2012} result.  Then, we reduced a subset
of the data to mimic the depth of the PHAT photometry in order to look
for biases in the distribution of ages younger than 5 Gyr that could
potentially be attributed to the shallower nature of the PHAT survey.
Even using only a small subset of the \citet{bernard2012} data
resulted in relatively deep photometry due to the low stellar density
of the far outer disk, so we limited the fitting to a portion of the
CMD similar to the portion we were able to fit using the PHAT data in
our inner fields (see black line in the upper-right panel of
Figure~\ref{comp_bernard}).  Thus, when fitting the shallow data, the
fitting routine had no access to the subgiant branch or lower
main-sequence, as these features fall outside of the included portion
of the shallow CMD.

Figure~\ref{comp_bernard} shows comparison between the age
distributions younger than 5 Gyr from fitting the full depth of the
data and from the much shallower subset of the data. We do not include
ages greater than 5 Gyr on the plots to focus on the performance in
the epoch of interest.\footnote{The CMDs all contain a significant
  fraction of stars older than 5 Gyr, which were also measured by the
  SFH fitting. We leave the contribution of these stars out of the
  current analysis.}  Thus the cumulative fractions are relative to
only the subset of stars with ages $<$5 Gyr, not the total stellar
population.  We outline the timing of the burst we find in the inner
disk by shading different epochs. The age distribution from the full
depth is shown with the dashed line.  As published in
\citet{bernard2012} this SFH has a lull in star formation at 4-5 Gyr
followed by a sharp burst at 2-3 Gyr.  The fit to the shallower data
also detects a burst at 2-3 Gyr, but ascribes $\sim$25\% more of the
$<$5 Gyr old stellar population to the burst than the fit to the deep
data. Thus, both measurements agree on the presence of the burst,
though its exact strength is more uncertain for the shallower data.
However, the two measurements are fully consistent within the
uncertainties, shown as the gray shaded region, confirming that our
technique accurately estimates the precision with which we can
determine the SFH in this age range.

\section{Results and Discussion}

\begin{figure}
\includegraphics[width=3.05in]{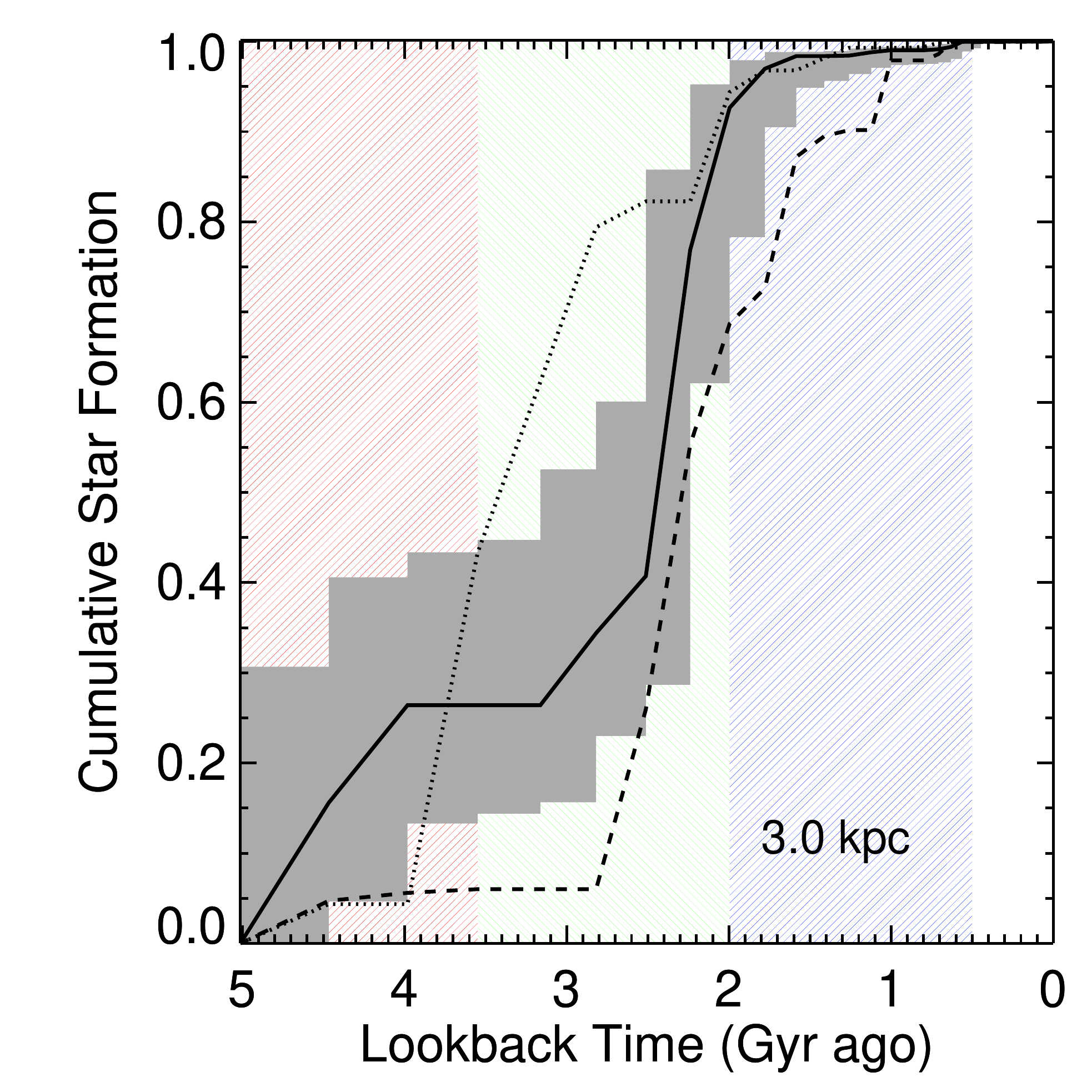}
\centerline{\hspace{-4.0in}\includegraphics[width=3.05in]{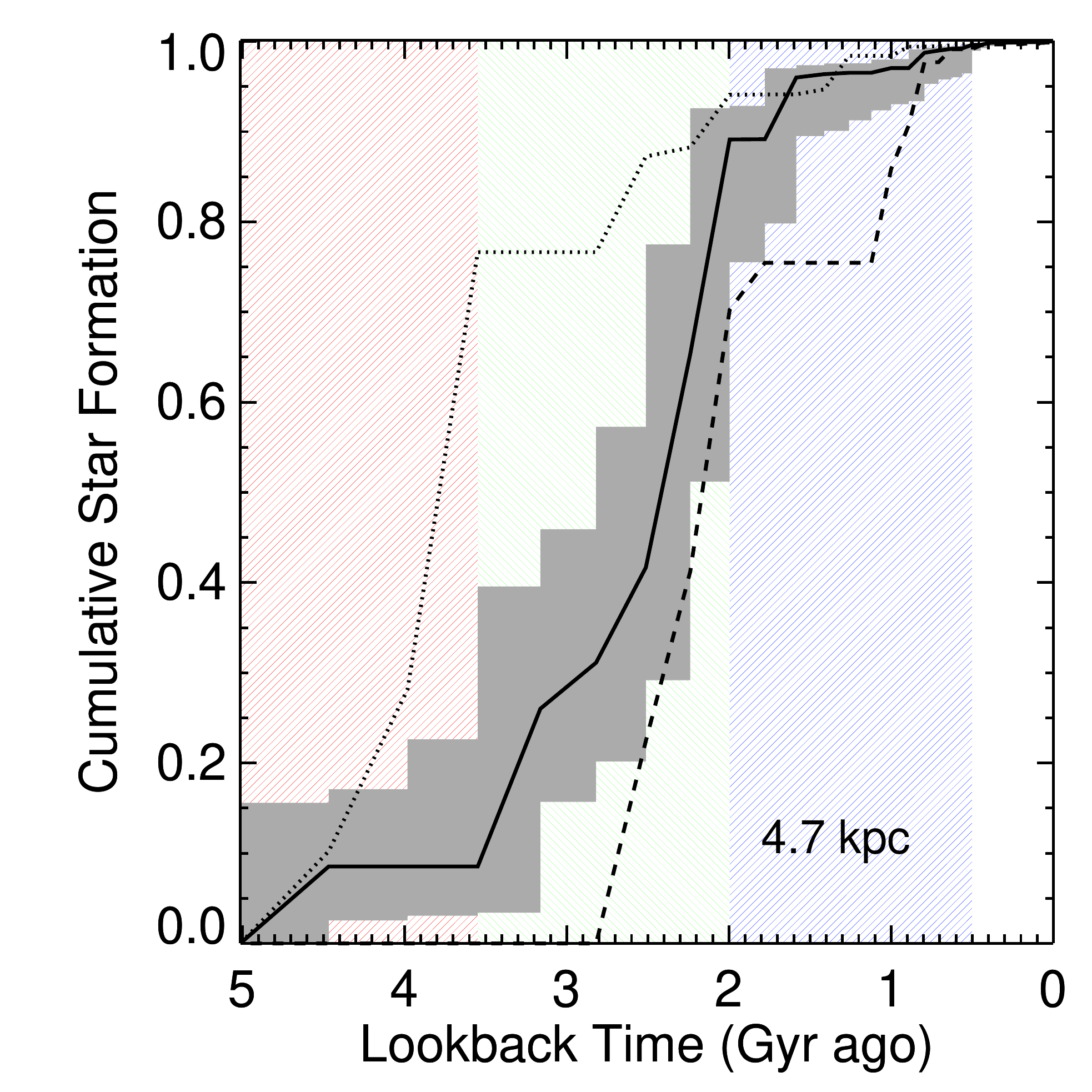}}
\includegraphics[width=3.05in]{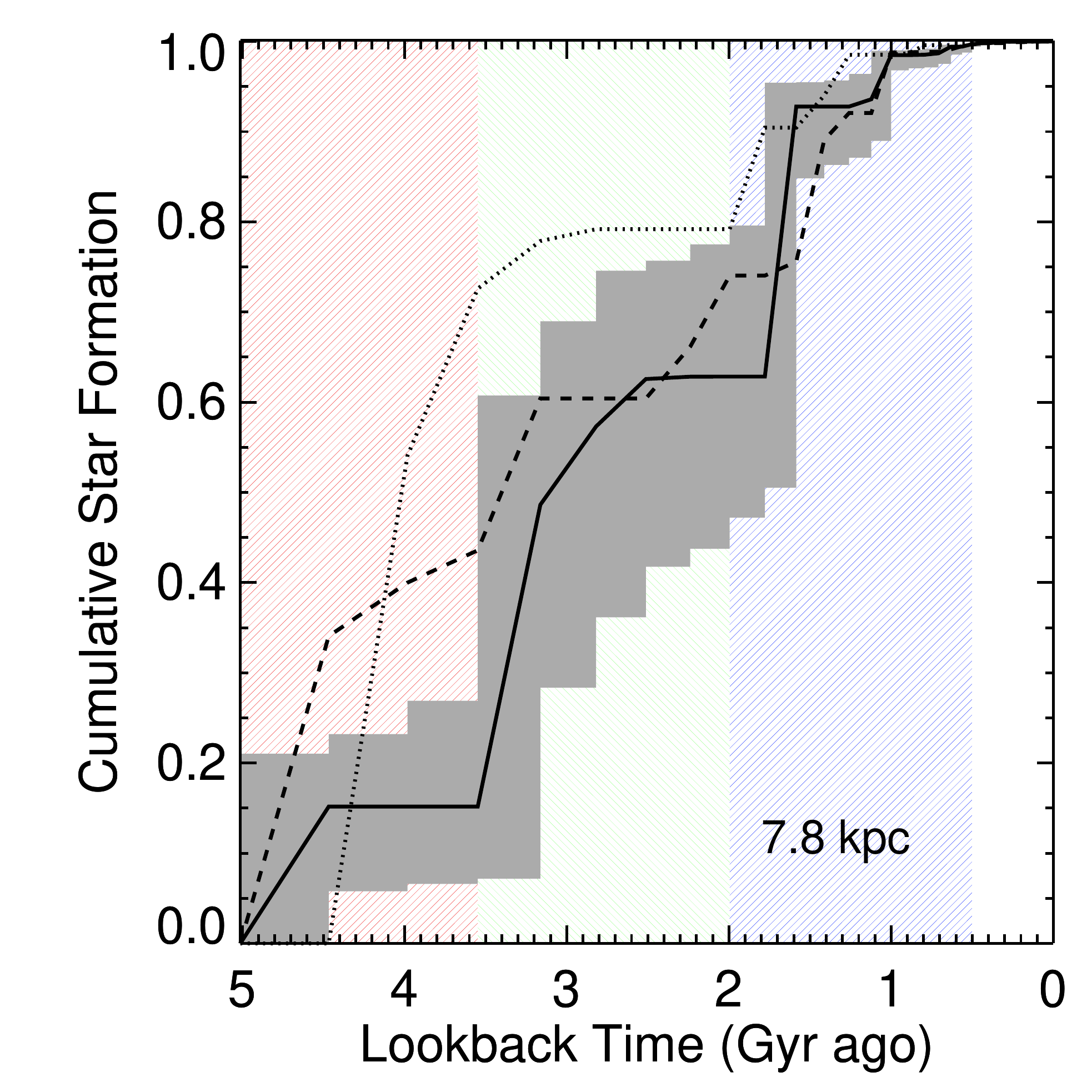}
\centerline{\hspace{-4.0in}\includegraphics[width=3.05in]{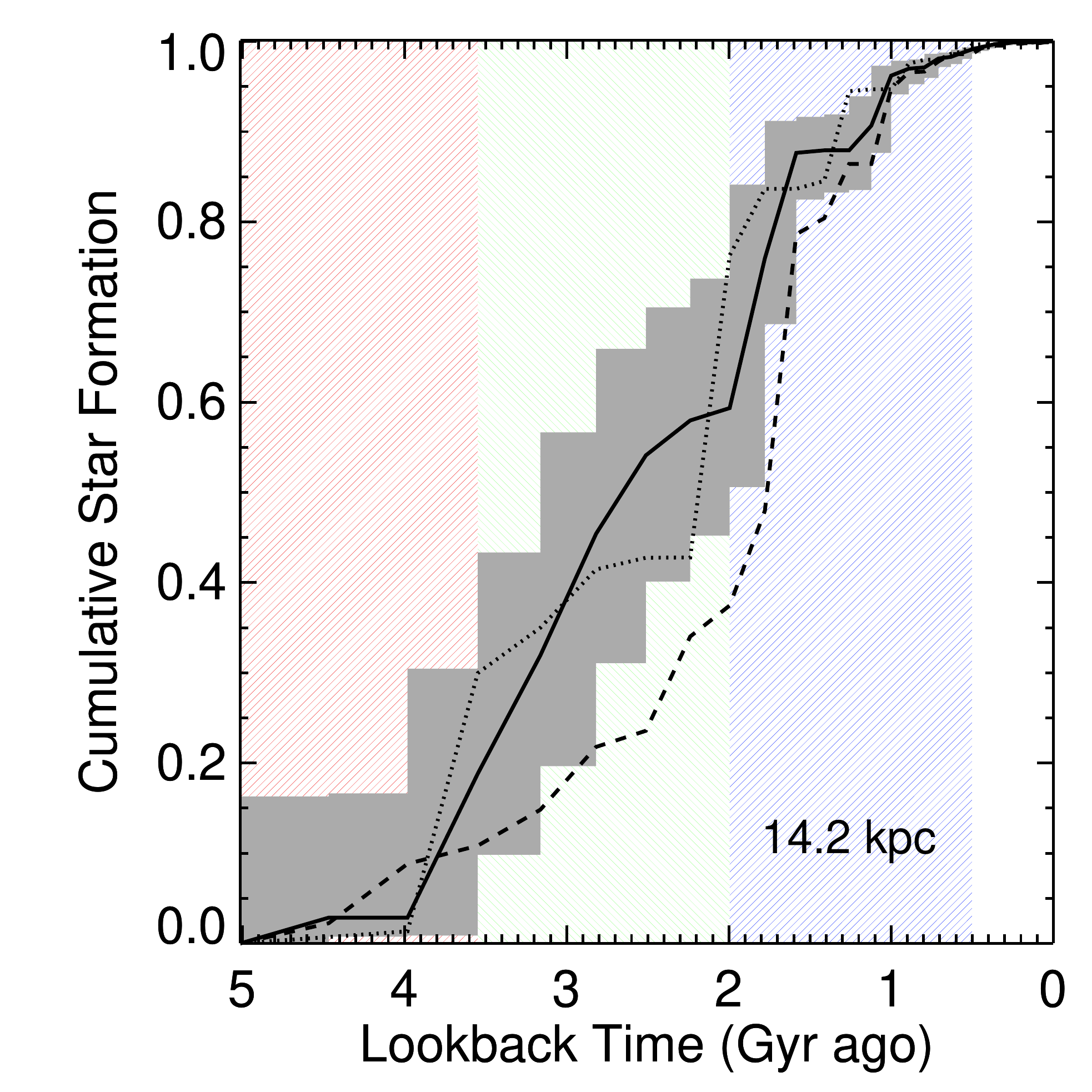}}
\includegraphics[width=3.05in]{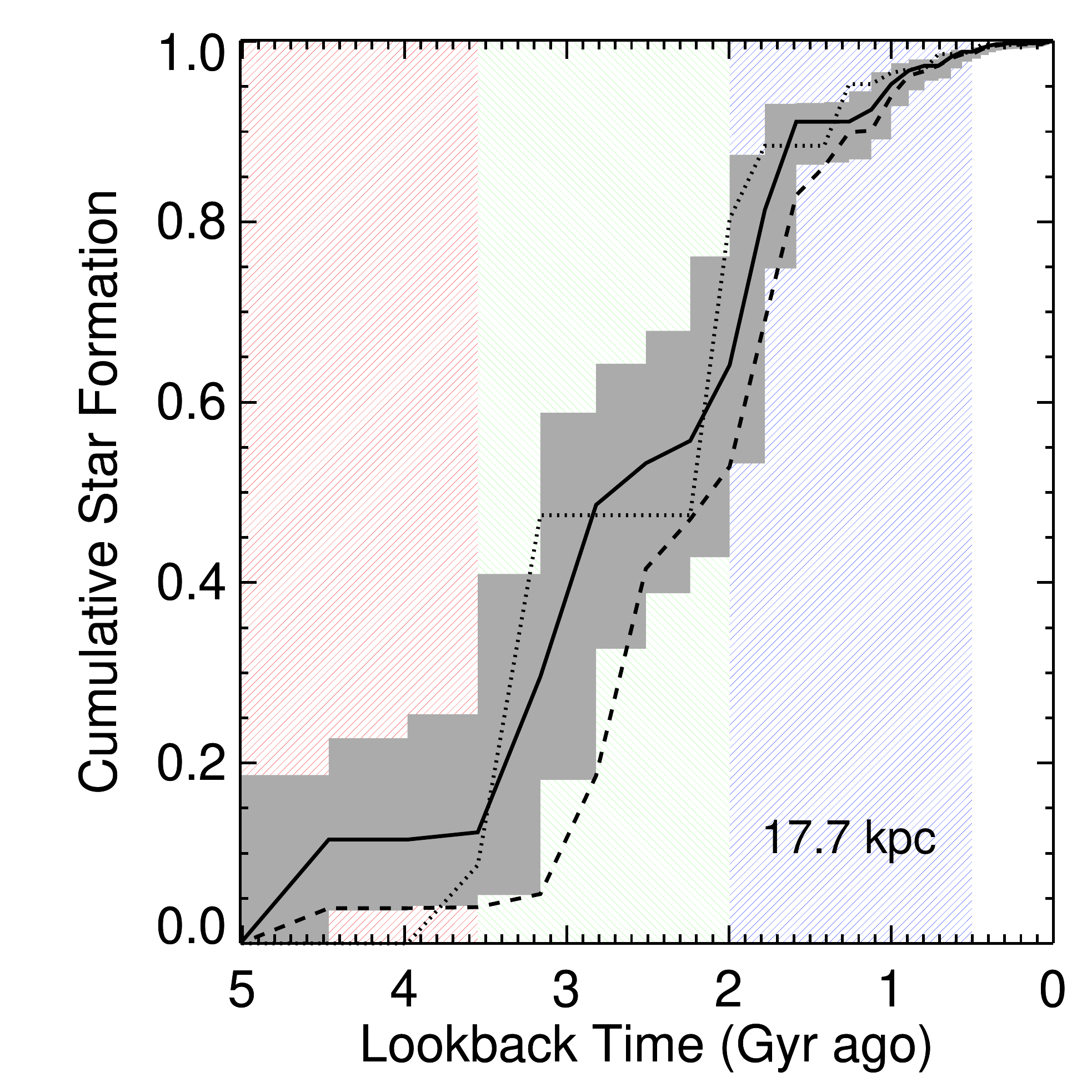}
\includegraphics[width=3.05in]{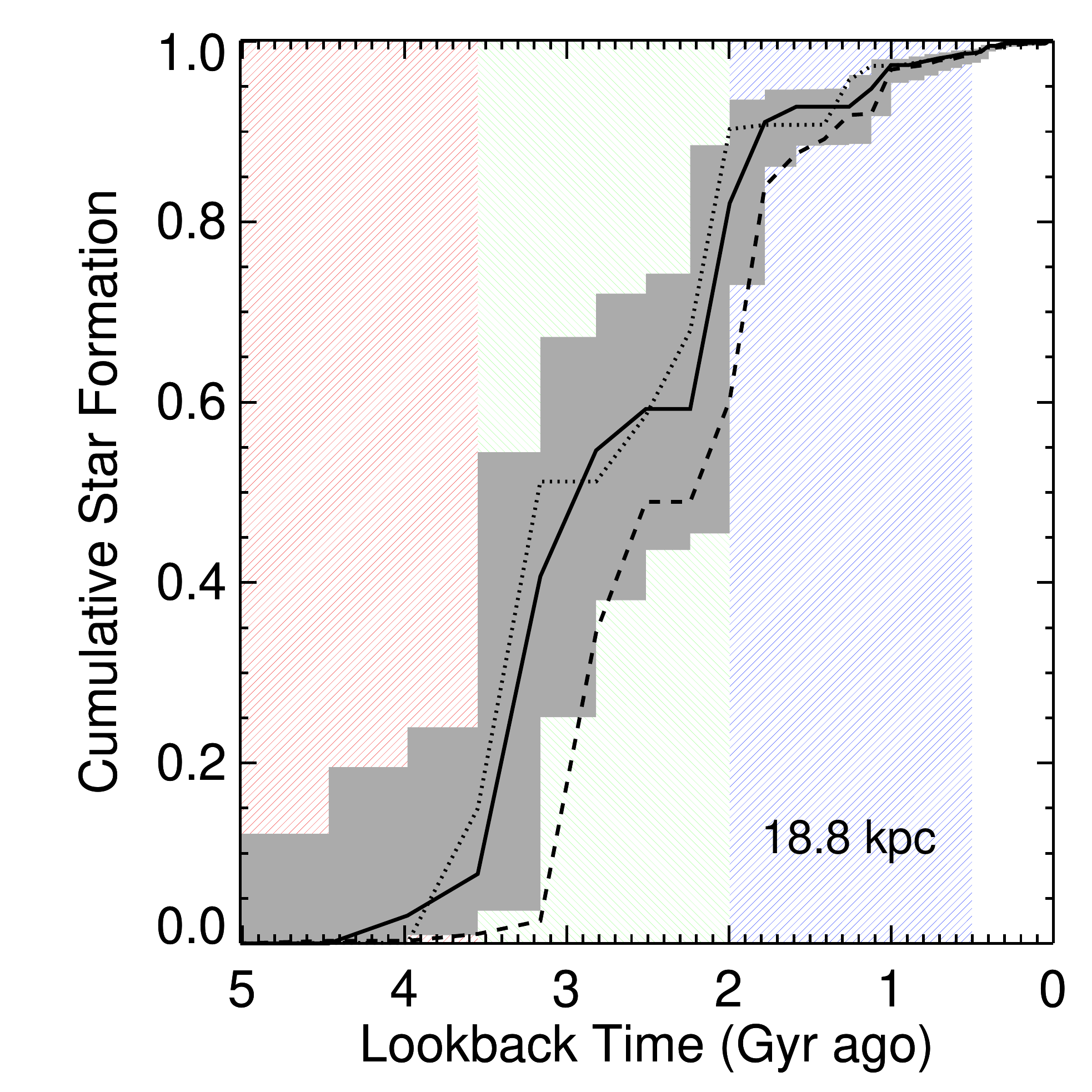}
\caption{The cumulative star formation for the past 5 Gyr at 6
  different radii. Other regions are not shown for brevity.  These
  correspond to the six radial groupings along the major axis shown in
  Figure~\ref{locations}.  Solid lines with gray shading show the fits
  to the Padova models along with the total uncertainties described in
  Section 2.1.  Dashed lines show the best fit to the BASTI models,
  and dotted lines show the best fit to the PARSEC models.  Three
  colored stripes show the 3 epochs used to make the plots in
  Figure~\ref{profiles}, in their respective colors. All of the model
  sets result in a steep increase of short duration in all locations.
  The Padova models measure the burst to be 2.0-3.5 Gyr ago (green
  area of the plots) relative to the neighboring epochs. As in
  Figure~\ref{comp_bernard}, the cumulative fractions refer only to
  those stars formed in the last 5 Gyr.}
\label{recent_sfhs}
\end{figure}

In Figure~\ref{recent_sfhs}, we show the cumulative age distributions
as a function of lookback time for six of our eight groups.  As in
Figure~\ref{comp_bernard}, we focus on the stars formed in the past 5
Gyr.  The uncertainties for our adopted fits (those with the Padova
models) are shown on the cumulative distributions as gray shaded
areas.  We also show the best fits from the BASTI and PARSEC models as
dashed and dotted lines, respectively.

\subsection{A Global Burst of Star Formation}

Fits to all model sets show that the population of stars with ages
$<$5 Gyr is dominated by a strong episode of star formation. All of
our SFHs show that star formation essentially shuts down over the past
1-2 Gyr, after undergoing a significant burst of star forming activity
2.0-3.5 Gyr ago.  The possible exception is in the group near 7.8 kpc,
where star formation was more steady.

\begin{figure}
\includegraphics[width=6.2in]{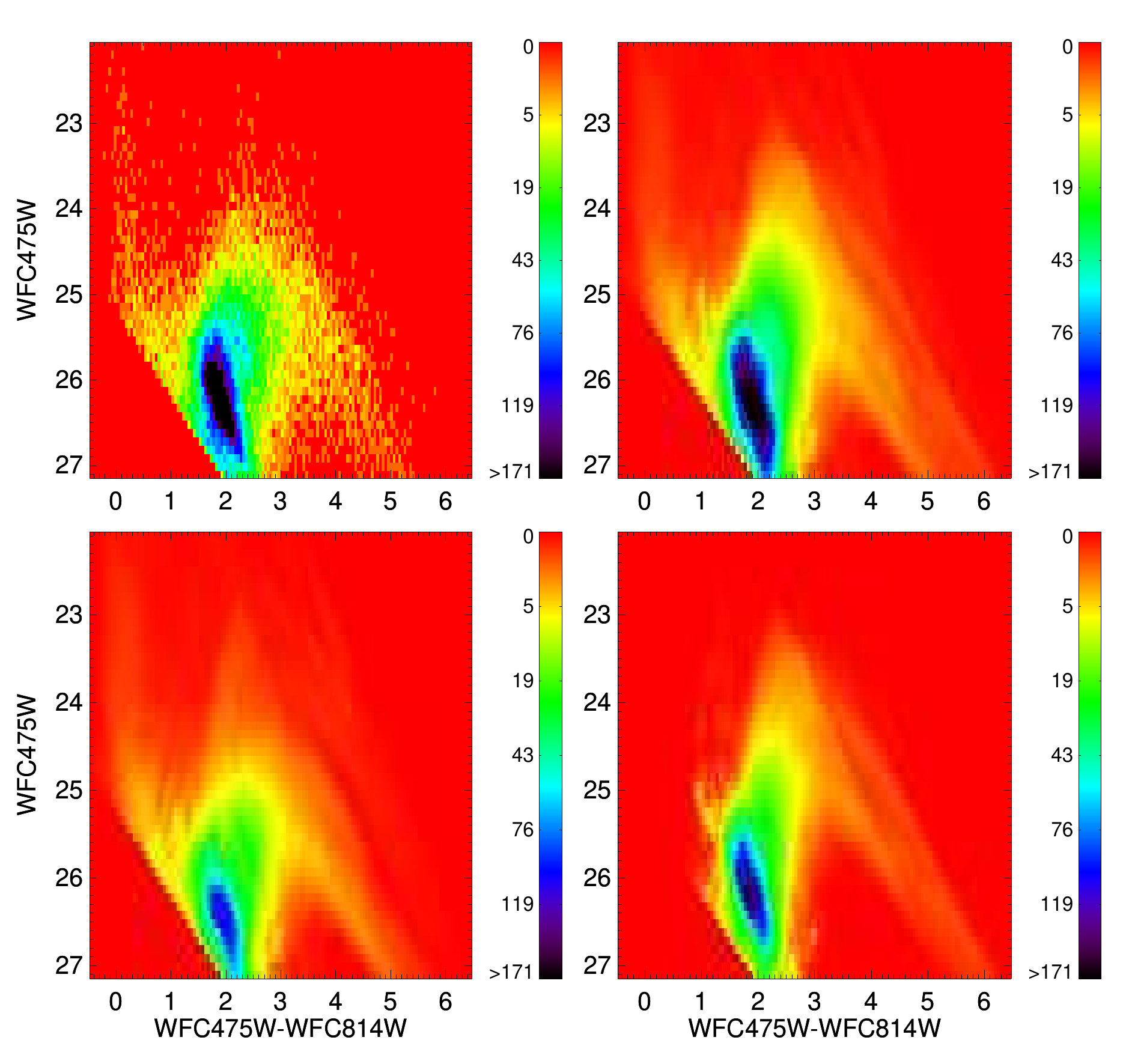}
\caption{{\it Upper Left:} Observed CMD for the 4.7 kpc group.  {\it Upper Right:} Best-fit Padova model CMD. {\it Lower Left:} Best-fit Padova model excluding all stars with ages from 2.0 to 3.5 Gyr. {\it Lower Right:} Difference between {\it upper right} and {\it lower left}, isolating the CMD features that indicate the presence of the enhancement at 2.0-3.5 Gyr.}   
\label{diff_models}
\end{figure}

Since our data are not of sufficient depth to probe the main sequence
turnoff for stars of ages 2-3 Gyr, it is not necessarily instantly
obvious which CMD feature is driving the detection of the large
population of stars with these ages in our data.  To qualitatively
assess the detection, we produced model CMDs using the best fit and
omitting the star formation from 2.0-3.5 Gyr, using the 4.7 kpc group
as a fiducial.  These model CMDs, along with the observed CMD and the
difference between the two, are shown in Figure~\ref{diff_models}.
The driving features appear to be the bright-blue portion of the red
clump, the asymptotic giant branch, and the relatively blue and
vertically oriented portion of the red giant branch.  These features
are apparently fit best by models of 2.0-3.5 Gyr in age.  We have
verified that models of older age with lower metallicity provide a
poorer fit, as do models of constant star formation rate.

We note that these CMD features are not the most well-tested in the
stellar evolution models \citep[e.g.][]{gallart2005}.  Therefore,
there is a chance that future changes to models of the red giant branch,
asymptotic giant branch, and red clump models and/or deeper data could
prove the population consistent with a more constant SFH.  However,
the detection of a burst of the same age in 14 deeper fields in the
outer disk (25-90 kpc deprojected radii) at a variety of azimuthal
angles \citep{bernard2015}, along with the agreement between our
fields of depths varying by 2 magnitudes and covering 17 kpc of the
inner disk in deprojected galactocentric distance, strongly suggests
that our result is robust given the models currently available for
fitting resolved stellar photometry.  

\subsubsection{Age of the Burst}

While the detection of a burst of star formation in M31 in the past 5
Gyr appears clear, it is difficult to pinpoint the age of the episode
given the depth of our data.  The age of the episode is most uncertain
at the inner radii where the data are shallower, as shown by both the
uncertainties and the very different ages measured from fits to
different models. We note that if we could isolate the age more
precisely than 1 Gyr, the burst could have been shorter and more
intense; thus, the measured amplitude of the increase is a lower limit
and the duration of $\sim$1.5 Gyr is an upper limit.  The model
dependence of the age means that our precision on the burst age is
limited to 2-4 Gyr ago (single digit precision) although our analysis
with the Padova models alone would place the age at 2.0-3.5 Gyr.

\subsubsection{Stellar Mass Produced}

To investigate the amount of stellar mass formed in this burst, we
calculated the surface density of the stellar mass formed from 2.0 to
3.5 Gyr ago as a function of median radius, and we compared this with
the surface density of stars formed over other epochs of similar
duration from 0.5-5.0 Gyr ago. These epochs are marked on
Figure~\ref{recent_sfhs}. We avoid the most recent 500 Myr as our
regions were chosen to avoid dust, making us biased against recent
star formation, and because this time period is investigated in detail
in \citet{lewis2015}. We plot these profiles in the left panel of
Figure~\ref{profiles}.  We note that the results from the shallower
data at small radii are consistent with an extrapolation of the
measurements from deeper data at larger radii.  The right panel shows
a sand-pile histogram of the fraction of the total 0.5-5 Gyr old
stellar mass formed in each epoch.

Integrating the exponential profile plotted for the 2.0-3.5 Gyr time
period shown in green in Figure~\ref{profiles} (see
Table~\ref{exponential_tab}), assuming azimuthal symmetry, yields
8.1$\times$10$^{9}$~M$_{\odot}$ of stars formed.  Integrating the
exponential profile for the 3.5-5.0 Gyr period yields only
2.5$\times$10$^{9}$~M$_{\odot}$, and the profile for the 0.5-2.0 Gyr
epoch also yields only 2.2$\times$10$^{9}$~M$_{\odot}$.  Thus, this
was a galaxy-wide factor of $\sim$3-4 increase in the star formation
rate averaged over 1.5 Gyr, resulting in the production of
6$\times$10$^{9}$~M$_{\odot}$ of additional stellar mass.  Assuming
$\sim$30\% star formation efficiency, this mass would be associated
with 2$\times$10$^{10}$~M$_{\odot}$ of baryonic matter.

This large amount of star formation likely significantly depleted the
gas supply, resulting in the low SFR seen today.  The current gas mass
in M31 is $\sim$7$\times$10$^{9}$~M$_{\odot}$ \citep{draine2014}, and
the star formation rate is $\sim$0.7~M$_{\odot}$~yr$^{-1}$, yielding a
depletion timescale of $\sim$10 Gyr.  However, the low SFR of M31 puts
it into a transition zone between the blue and red bimodal color
populations of large galaxy surveys \citep{mutch2011}.  Thus, this
recent burst may mark an event in M31 history that triggered the
current global transformation toward the red sequence.  We consider
whether this star formation 2.0-3.5 Gyr ago may signify an interaction
or merger below.

\section{Cause of the Burst}

There are at least 3 possible scenarios that could induce such global
star formation in M31. One possibility was suggested by
\citet{bernard2015}: a tidal interaction with M33 during that epoch.
Another possibility is a very strong interaction with M32, which
stripped M32 into the dwarf elliptical that it is today.  Finally,
there is the possibility that a relatively major merger coalesced 2
Gyr ago into what is now M31.  We now discuss each of these
possibilities in turn.

\subsection{Interaction with M33 or M32}

It is possible that an interaction with another Local Group galaxy
triggered the star formation episode in M31 2-4 Gyr ago.  If so,
based on previous studies, the most likely candidates are M33 and/or M32.

The timing of the burst at 2-3 Gyr ago is consistent with that seen in
the outer M31 disk by \citet{bernard2015} with deeper data over 14
fields spread all over the outer M31 disk.  Simulations that reproduce
the current velocities and positions of M31 and M33 suggest a close
passage 2-3 Gyr ago\citep{mcconnachie2009}.  Furthermore, a similar
burst of similar age is seen in the star formation history of M33
\citep[see SFHs in][]{williams2009}.  Numerical simulations do show
that interactions can increase star formation rates by factors of 4-5
\citep[e.g.][]{springel2000}, but such increases are only seen in
simulations of very close interactions between galaxies of similar
mass.

There are some characteristics of the M31-M33 system that make the
M31-M33 interaction explanation for such a star formation episode less
clear.  First is the amount of mass in stars formed.  Second is galaxy
mass and morphology.

Our measurements suggest that this burst was widespread and produced a
large amount of stellar mass in the massive galaxy M31.  A fly-by of a
galaxy $\sim$10\% of the mass of M31 may not be provide enough of a
perturbation and may not include enough gas to explain this amount of
increased star formation.

Furthermore, the masses and morphologies of the two galaxies do not
seem consistent with a strong interaction between them.  When smaller
galaxies interact with larger galaxies, typically the morphology of
the less massive galaxy is more affected than that of the more massive
one \citep[e.g.,][]{mayer2001,governato2004}.  M33 has a relatively
undisturbed morphology, showing only a warped outer disk and a disk
break beyond 8 kpc \citep{ferguson2007,williams2009,mcconnachie2009}
despite being significantly less massive.  M31 has a highly evolved
morphology, as if it has been subjected to strong interactions,
forming a large bulge and relatively quiescent disk.  Therefore, the
morphologies of the two galaxies (especially that of M33) are not
those expected if they have recently been through a strong
interaction.

Still, the simultaneous nature of the episodes seen in M31 and M33
suggest that there could have been an interaction, and the simulations
of \citet{mcconnachie2009} suggest that such an interaction occurred
$\sim$2-3 Gyr ago.  Perhaps future simulations will show that a major
star formation episode can be triggered in a massive galaxy without
causing major morphological changes to a low-mass galaxy.  Perhaps
such a burst could occur if the M31 disk was relatively gas-rich at
the time and if the interaction had just the right impact parameter
and orientation to induce some kind of resonance in the disk.
Detailed hydrodynamic simulations of all possible interactions between
the systems may be required to answer this question.

On the other hand, the M31-M32 system appears more consistent with
having undergone a strong interaction a few Gyr ago.
\citet{block2006} simulate a collision between M31 and M32 $\sim$200
Myr ago, which is more recent than our measurements probe, but
suggests that the system has interacted strongly in the past.
Morphologically, as a very compact dwarf elliptical very close to M31,
M32 appears more likely to have been strongly influenced by an
interaction with M31 in the recent past. Furthermore, there is a
significant 2-5 Gyr old population in M32 which comprises 40\% of its
mass \citep{davidge2007,monachesi2012}.  It is possible that M32 is
the remnant of a galaxy that was once more massive and gas rich
\citep{bekki2001}. This relatively gas rich galaxy could have strongly
interacted with M31 about 2 or 4 Gyr ago, triggering the burst in M31
as well as the centrally-concentrated star formation in M32,
generating the 2-5 Gyr old population in M32, and leaving M32 the
compact dwarf elliptical galaxy we see today. On the other hand, M32
does not currently show kinematic or structural evidence of a strong
tidal disturbance \citep{lauer1998,howley2013}

There is also the possibility that both of these scenarios occurred
during the epoch in question.  Perhaps M33 was much closer to M31
during this epoch (as in the simulations of
\citealp{mcconnachie2009}), explaining the increase SFR in M33 at that
time.  Furthermore, M32 could have been in the process of being
stripped and both events could have contributed to the elevated star
formation in M31.

\subsection{Possible Merger} 

Another possible explanation for the episode is a merger with galaxy
$\gap$20\% of the pre-merger mass of M31, such as in the merger
simulations of \citet{cox2008}. The density profile of the stars
formed, the recent decrease in the M31 star formation rate, and the
stellar mass produced are all consistent with a relatively large
merger scenario, as described below.

\begin{figure}
\includegraphics[width=3.5in]{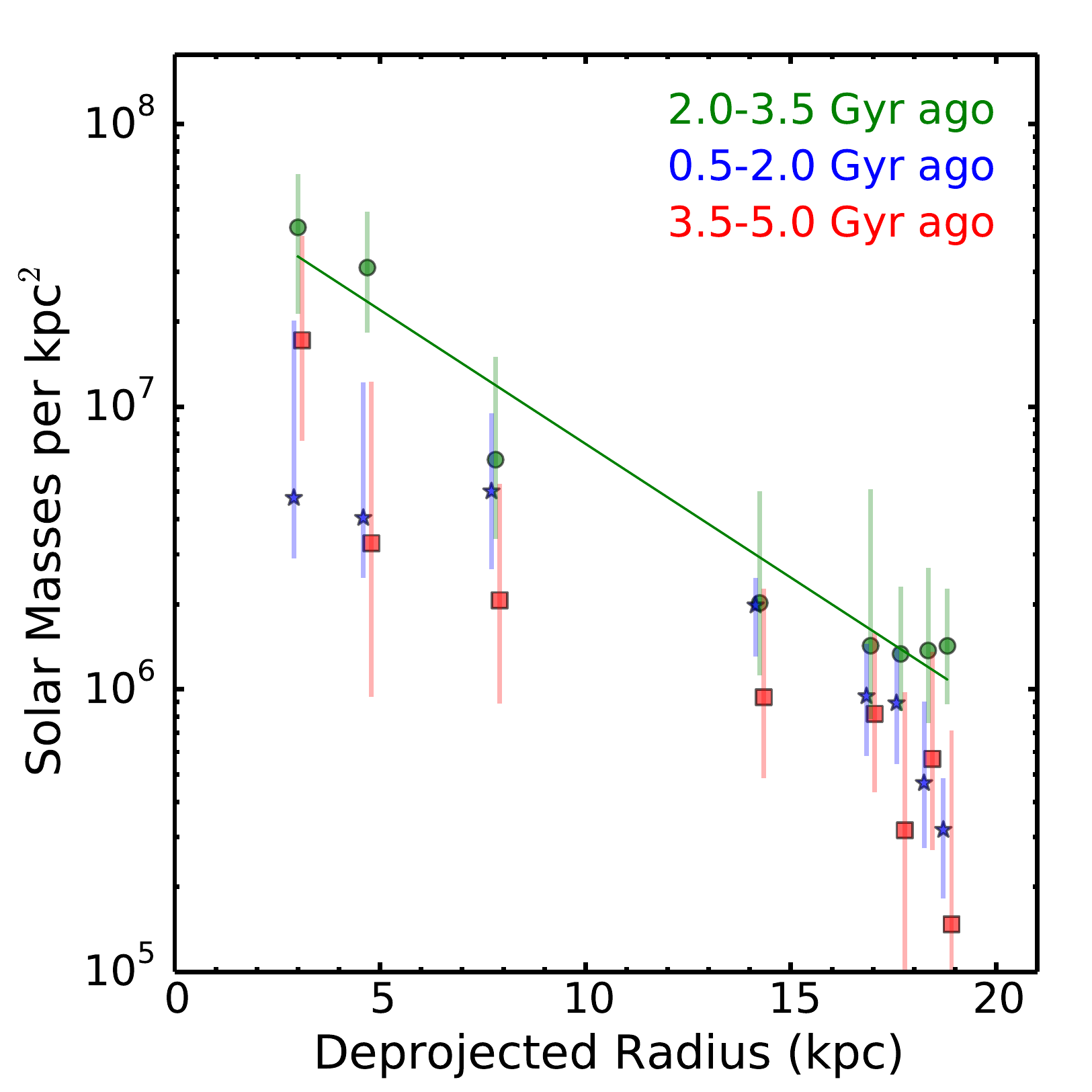}
\includegraphics[width=3.5in]{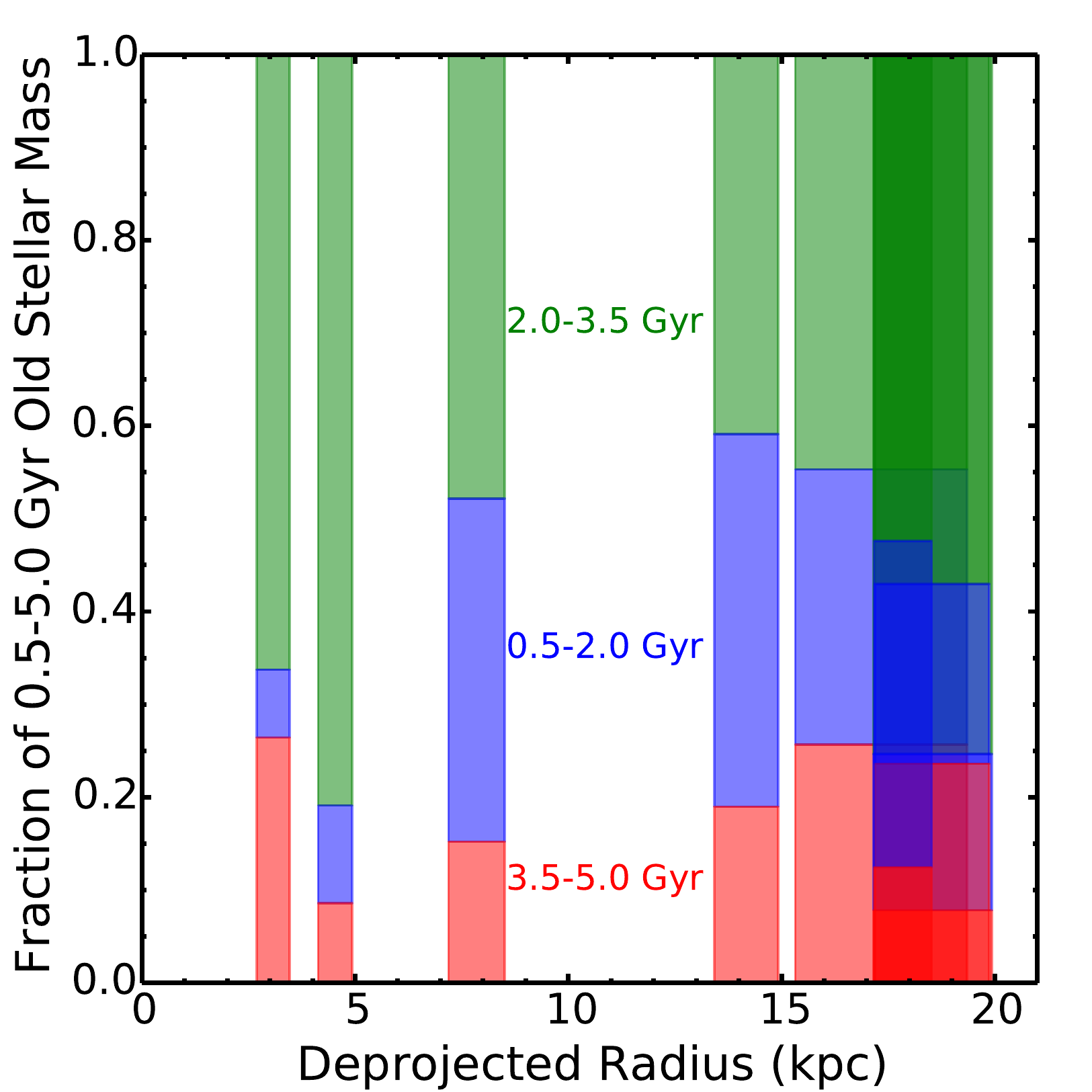}
\caption{{\it Left:} Radial profiles of the stellar mass surface
  density of M31 in three age bins (see Table~\ref{exponential_tab}).
  Colors and symbol types represent the stellar mass contribution of
  several epochs of similar duration, shown as colored vertical
  stripes in Figure~\ref{recent_sfhs}.  The 2.0-3.5 Gyr old population
  (green/circles) dominates the stellar surface density.  The best-fit
  exponential to this population is overplotted with the green line
  (6.6$\times$10$^7e^{-r/4.6 kpc}$). {\it Right:} Sand-pile histogram
  of the fraction of the 0.5-5.0 Gyr old population in each epoch is
  plotted as a function of deprojected radius.}
\label{profiles}
\end{figure}

Like the \citet{cox2008} simulations, the density profile of the
enhancement is similar in shape to the overall surface density profile
of M31, as shown in Figure~\ref{profiles}.  Furthermore, the
simulations show a smooth decrease in global star formation rate down
to levels $\lap$1 M$_{\odot}$ yr$^{-1}$ in the 2 Gyr following the
merger as a result of gas depletion.  Such a decrease is consistent
with that recently measured by \citet{lewis2015}.  The stellar density
in the outer regions of merger simulations
\citep[e.g.,][]{cox2008,moreno2015} $\sim$3 Gyr after the merger show
only relatively faint structures, qualitatively similar to those
observed in the M31 halo
\citep{ibata2001,ferguson2002,mcconnachie2009}.

The amount of stellar mass produced by the event hints at a relatively
large merger.  If the matter that formed the stars in the burst came
from the galaxy that merged, the baryonic Tully-Fisher relation
\citep[e.g.]{zaritsky2014} yields a circular velocity of $\sim$140 km
s$^{-1}$ for 2$\times$10$^{10}$~M$_{\odot}$ of baryonic mass, which is
similar to the circular velocity of M33 \citep{corbelli2000}. As some
of the mass was in M31 already, this estimate is an approximate
upper-limit on any infalling galaxy.

%Thus, these measurements suggest just a few Gyr
%ago, the third or fourth most massive member of the Local Group merged
%with M31.

%converts to $\sim$5$\times$10$^{11}$~M$_{\odot}$ of total
%mass if calibrated to the Milky Way.  However, this is an upper limit
%to the accreted mass, since M31 likely used most of its own gas during
%this burst as well.  If we assume half of the additional mass came
%from the accreted galaxy, the circular velocity of the accreted mass
%would be $\sim$110 km s$^{-1}$, which would correspond to a total mass
%of $\sim$3$\times$10$^{11}$~M$_{\odot}$.  Thus, these measurements
%suggest just a few Gyr ago, the third or fourth most massive member of
%the Local Group merged with M31.

It is difficult to compare our SFHs of relatively low time resolution
to the SFHs of the \citet{cox2008} simulations directly. However, a
1:1 merger ratio in their simulations appears to produce more than the
equivalent of a factor of 2 increase in star formation rate for 1.5
Gyr.  Such an equal-mass merger appears to produce the equivalent of
about a factor of $\gap$10 increase for 1 Gyr.  The 5.8:1 merger does
not appear to enhance the star formation enough to produce the effect
we see in the M31 data.  The mean rate increase of a factor of 4-5
over about a Gyr we see in our data is limited by our time resolution.
The actual burst could have had a much shorter duration, but produced
enough stellar mass for us to detect it throughout the galaxy.  Most
consistent with merger simulations would be another galaxy $\gap$20\%
of the pre-merger mass of M31. Such a total mass
($\sim$2$\times$10$^{11}$ M$_{\odot}$) is reasonable considering the
baryonic mass estimate above of $\sim$2$\times$10$^{10}$~M$_{\odot}$.
 
The merger scenario also has problems.  First, it is unclear if the
M31 disk would have survived, even in its relatively quiescent state,
if it had undergone such a merger.  However, simulations do suggest
that disks can survive large mergers if they contain high gas
fractions \citep{hopkins2009}.  Second, it is puzzling why M33 would
show a simultaneous global star formation episode.  However, it is
possible that the galaxy that merged with M31 passed by M33 (which may
have been much closer to M31 at the time \citealp{mcconnachie2009})
along the way, inducing star formation during the same epoch but not
strongly disturbing its morphology.  Finally, very recent simulations
by \citet{moreno2015} suggest that most of the star formation in
mergers occurs $<$1 kpc from the galaxy center, which would not be
consistent with this disk-wide burst in M31.

Whatever the explanation, these newly-measured age distributions from
the PHAT survey provide a new piece of evidence, along with the
structured halo \citep{hammer2010,hammer2013,sadoun2014}, structured
satellite distribution \citep{ibata2013,ibata2014a}, peak in massive
cluster ages \citep{fusipecci2005}, and velocity dispersion of the
disk \citep{dorman2015} that point to M31 having undergone a
significant interaction event in the past few Gyr.

In conclusion, we have found new evidence that a global burst of star
formation occurred in M31 2-4 Gyr ago over the entire M31 disk.  The
cause of this burst is far from clear.  It could have been due to
interactions with M32 and/or M33.  It could have been a merger, or it
could be some combination of these.  The occurrence of the burst may
help to explain the current, low-activity state of the M31 disk and
its current global transformation toward the red sequence.  Future
simulations may be able to provide more detailed constraints on the
physical cause of the burst.

\bigskip 

Support for this work was provided by NASA through grant GO-12055 from
the Space Telescope Science Institute, which is operated by the
Association of Universities for Research in Astronomy, Incorporated,
under NASA contract NAS5-26555. Support for DRW is provided by NASA
through Hubble Fellowship grants HST-HF-51331.01 awarded by the Space
Telescope Science Institute.  We thank Amazon cloud services, for
donating some of the computing time necessary to make these
measurements.

%\bibliographystyle{apj}
%\bibliography{apjmnemonic,event} 

%\clearpage
\FloatBarrier
\begin{deluxetable}{cccccccc}[H]%\tablewidth{18cm}
\tablecaption{Properties of the 9 samples taken from the PHAT dust-free regions}
\tablehead{
\colhead{Region} & 
\colhead{R$_{deproj}$ (kpc)} &
\colhead{ Min$_{deproj}$ (kpc)} & 
\colhead{Max$_{deproj}$ (kpc)} & 
\colhead{Area (kpc$^2$)} & 
\colhead{N$_{\rm stars}$} & 
\colhead{$\Sigma$ (arcsec$^{-2}$)} &
\colhead{d$A_V$ (mag)} 
}
\startdata
%1 & 1.49 & 1.04 & 1.81 & 0.078 & 116980 & 0.6\\
1 & 3.00 & 2.69 & 3.46 & 0.075 & 122561 & 22.7 & 0.4\\
2 & 4.69 & 4.13 & 4.93 & 0.034 & 50177 & 20.3 & 0.6\\
3 & 7.81 & 7.19 & 8.51 & 0.182 & 198732 & 15.2 & 0.4\\
4 & 14.24 & 13.42 & 14.92 & 0.075 & 47038 & 8.7 & 0.2\\
5 & 16.94 & 15.32 & 19.35 & 0.245 & 131251 & 7.5 &0.2\\
6 & 17.67 & 17.17 & 18.52 & 0.044 & 15078 & 4.8 & 0.0\\
7 & 18.34 & 17.19 & 19.87 & 0.288 & 91090 & 4.3 & 0.0\\
8 & 18.81 & 17.15 & 19.93 & 0.191 & 48055 & 3.5 & 0.2
\enddata
\label{locations_tab}
\end{deluxetable}

\begin{deluxetable}{cccc}[H]%\tablewidth{18cm}
\tablecaption{Exponential profile fit parameters and integrated stellar masses for the three age ranges}
\tablehead{
\colhead{Age Range} &
\colhead{Scale Length (kpc)} &
\colhead{Normalization (M$_{\odot}$ kpc$^{-2}$)} &
\colhead{Integrated Stellar Mass (M$_{\odot}$)} 
}
0.5-2.0 Gyr & 7.1$\pm$1.0 & 8.9${\pm}3.7{\times}10^{6}$ &  2.2${\pm}1.1{\times}$10$^9$\\ 
2.0-3.5 Gyr & 4.5$\pm$0.5 & 6.6${\pm}2.1{\times}10^{7}$ & 8.0${\pm}2.5{\times}$10$^9$\\
3.5-5.0 Gyr & 4.9$\pm$0.8 & 1.8${\pm}1.0{\times}10^{7}$ & 2.5${\pm}1.1{\times}$10$^9$
\enddata
\label{exponential_tab}
\end{deluxetable}

%\begin{figure}
%\includegraphics[width=3.0in]{f5a.pdf}
%\centerline{\hspace{-4.0in}\includegraphics[width=3.0in]{f5b.pdf}}
%\includegraphics[width=3.0in]{f5c.pdf}
%\centerline{\hspace{-4.0in}\includegraphics[width=3.0in]{f5d.pdf}}
%\includegraphics[width=3.0in]{f5e.pdf}
%\includegraphics[width=3.0in]{f5f.pdf}
%\caption{The star formation surface density history for the past 5 Gyr
%  at 6 different radii along the major axis.  These correspond to the
%  six groupings shown in Figure~\ref{locations} and the six cumulative
%  SFHs plotted in Figure~\ref{recent_sfhs}. Note that the plot ranges
%  are substantially different, as the star formation surface density
%  is a strong function of galactocentric distance.}
%\label{recent_difs}
%\end{figure}

\end{document}